\documentclass[reprint,amsmath,amssymb,aps]{revtex4-2}
\usepackage{graphicx}
\usepackage{dcolumn}
\usepackage{bm}
\usepackage{float}
\usepackage{color}
\usepackage[dvipsnames]{xcolor}
\usepackage{hyperref}
\usepackage{graphicx}
\usepackage{braket}
\usepackage{appendix}
\usepackage{chemmacros}
\usepackage{natbib}
\usepackage{siunitx}
\usepackage{booktabs}
\hypersetup{colorlinks=true, citecolor=blue, urlcolor=blue, linkcolor=blue}

\begin{document}

\preprint{APS/123-QED}

\title{Breakdown of $J_{eff}$ = 0 and $J_{eff}$ = 3/2 states and existence of large magnetic anisotropy energy in vacancy ordered 5$d$ antifluorites: K$_2$ReCl$_6$, K$_2$OsCl$_6$, and K$_2$IrCl$_6$}

\author{Amit Chauhan}
\email{amitchauhan453@gmail.com}
\author{B. R. K. Nanda}
\email{nandab@iitm.ac.in}
\affiliation{$^1$Condensed Matter Theory and Computational Lab, Department of Physics, IIT Madras, Chennai-36, India}
\affiliation{$^2$Center for Atomistic Modelling and Materials Design, IIT Madras, Chennai-36, India}
\affiliation{$^3$Functional Oxides Research Group, IIT Madras, Chennai-36, India}
\date{\today}
\begin{abstract}
Vacancy-ordered antifluorite materials (A$_2$BX$_6$) are garnering renewed attention as novel magnetic states driven by spin-orbit coupling (SOC) can be realized in them. In this work, by pursuing density functional theory calculations and model studies, we analyze the ground state electronic and magnetic structure of face-centered cubic (fcc) antifluorites K$_2$ReCl$_6$ (KReC, 5$d^3$), K$_2$OsCl$_6$ (KOsC, 5$d^4$), and K$_2$IrCl$_6$ (KIrC, 5$d^5$). We find that KReC stabilizes in the high-spin $S$ = 3/2 state instead of the expected pseudo-spin $J_{eff}$ = 3/2 state. The former occurs due to large exchange-splitting as compared to the SOC strength. On the contrary, the KOsC stabilizes in broken $J_{eff}$ = 0 ($S$ = 1) simple Mott insulating state while KIrC stabilizes in $J_{eff}$ = 1/2 spin-orbit-assisted Mott insulating state. The presence of an isolated metal-chloride octahedron makes these antifluorites weakly coupled magnetic systems with the nearest and next-nearest-neighbor spin-exchange parameters ($J_1$ and $J_2$) are of the order of 1 meV. For KReC and KOsC, the $J_1$ and $J_2$ are estimated to be antiferromagnetic and ferromagnetic, which leads to a Type-I antiferromagnetic ground state, whereas for KIrC, both $J_1$ and $J_2$ are antiferromagnetic, hence, it stabilizes with a Type-III antiferromagnetic state. Interestingly, in their equilibrium structure, these antifluorites possess large magnetic anisotropy energy (0.6-4 meV/transition metal), which is at least one-to-two orders higher than traditional MAE materials like transition metals and multilayers formed out of them. Moreover, with epitaxial tensile/compressive strain, the MAE enhances by one order, becoming giant for KOsC (20-40 meV/Os). 
\end{abstract}
\maketitle

\section{Introduction}
Spin-orbit coupling (SOC) driven entanglement of spin and orbital degrees of freedom has led to the formation of novel quantum phases in transition-metal (TM) compounds involving chemically active 5$d$ valence electrons. Recently, TM compounds with heavier $d$ block elements such as Re, Os, and Ir have emerged as potential candidates due to the realization of topological phases such as Dirac and Weyl semimetals \cite{Ueda2018, Chauhan2022, Fujioka2019, Song2020}, spin liquid \cite{Okamoto2007, Kenney2019, Takahashi2019}, spin-orbit-assisted Mott insulator \cite{Pesin2010}, etc. in them. Apart from exploring the aforementioned phases, the current thrust in the research on SOC active 5$d$ quantum materials is to manifest the following two aspects.

\begin{figure}
\centering
\includegraphics[angle=-0.0,origin=c,height=8cm,width=8.5cm]{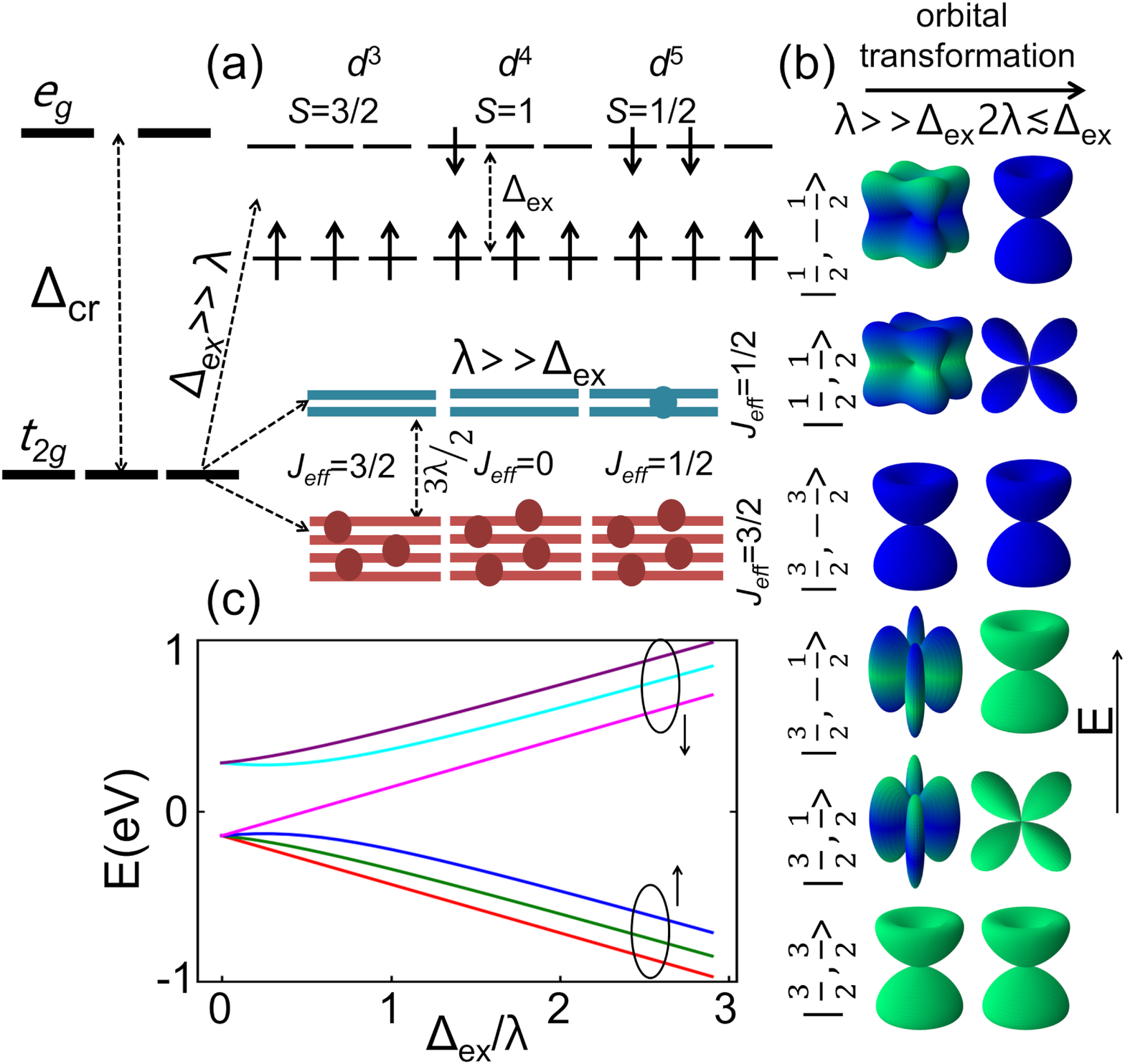}
\caption{(a) The splitting of five-fold degenerate $d$ manifold in an octahedral crystal field environment leading to three-fold degenerate $t_{2g}$ and two-fold degenerate $e_g$ states. With strong SOC and weak exchange-splitting ($\lambda$ $>>$ $\Delta_{ex}$), the $t_{2g}$ states further split to form four-fold and two-fold degenerate spin-orbital entangled $J_{eff}$ = 3/2 and $J_{eff}$ = 1/2 pseudo-spin states. In $\Delta_{ex}$ $>>$ $\lambda$ limit while the high-, intermediate-, and low-spin configurations stabilizes for $d^3$, $d^4$, and $d^5$ valence states, in the $\Delta_{ex}$ $<<$ $\lambda$ limit, $J_{eff}$ = 3/2, $J_{eff}$ = 0, and $J_{eff}$ = 1/2 pseudo-spin ground state forms. (b) The transformation of $J_{eff}$ = 3/2 and $J_{eff}$ = 1/2 states with $\Delta_{ex}$. The green and blue colors represent the spin-up and spin-down weights, respectively. (c) The energy eigen spectra as a function of $\Delta_{ex}$/$\lambda$ in the atomic limit. Increasing strength of exchange-splitting leads to the complete decoupling of spin-up and spin-dn channels. The $\lambda$ is chosen to be 0.3 eV which is the SOC strength of KReC.}
\label{orbitals}
\end{figure}

Firstly, the competition among crystal field splitting ($\Delta_{cr}$), onsite Coulomb repulsion ($U$), and SOC ($\lambda$) in an octahedral complex can lead to the formation of an intriguing electronic and magnetic structure in the pseudo-spin space. In an octahedral crystal field environment, the five-fold degenerate $d$ manifold splits into three-fold degenerate $t_{2g}$ and two-fold degenerate $e_g$ states (see Fig. \ref{orbitals}). With strong SOC, while the latter remains unperturbed, the former further splits into spin-orbital entangled $J_{eff}$ = 3/2 quartet (m$_j$ = $\pm 3/2, \pm 1/2$) and $J_{eff}$ = 1/2 doublet (m$_j$ = $\pm1/2$) with both of them separating energetically by 3/2 $\lambda$. For $5d^3$ electronic configuration, three electrons will occupy the lower-lying $J_{eff}$ = 3/2 states which will give rise to a weakly spin-polarized magnetic state due to equal and opposite spin/orbital polarization of each $m_j$ = $\pm 1/2$ or $\pm 3/2$ states.

Resonant inelastic X-ray scattering (RIXS) studies \cite{Taylor2017} on face-centered cubic (fcc) double perovskite compounds, Ca$_3$LiOsO$_6$ and Ba$_2$YOsO$_6$, reveal the stabilization of $J_{eff}$ = 3/2 electronic ground state, however, Ir based double perovskite Ba$_2$NiIrO$_6$ and Os based pyrochlore member Cd$_2$Os$_2$O$_7$ were found to stabilize in a $S = 3/2$ state rather than $J_{eff}$ = 3/2 state \cite{Calder2016, Yang2022}. Recently, the $d^4$ family of compounds are intensely explored to realize the $J_{eff}$ = 0 non-magnetic ground state due to fully occupied $J_{eff}$ = 3/2 manifold (see Fig. \ref{orbitals}) \cite{Nag2019, Bandyopadhyay2022, Song2022}. However, contrary to the expected ground state, recent RIXS and ab-initio studies report magnetic ground state for 6$H$-hexagonal perovskite iridate Ba$_3$ZnIr$_2$O$_9$, double perovskite iridate Sr$_2$YIrO$_6$, and quasi-one-dimensional spin chain system Sr$_3$NaIrO$_6$ to be magnetic \cite{Bandyopadhyay2022, Nag2019, Bhowal2015}. Though there is a lack of literature, the $d^5$ configuration in an ideal octahedron is bound and expected to show a single hole $J_{eff}$ = 1/2 state.

Secondly, from the application point of view, since the 5$d$ TM compounds are expected to demonstrate SOC-led anisotropy, there is a growing interest in calculating the magnetic anisotropy energy (MAE) in them \cite{Lau2019,Wang2017,Souza2022}. Large uniaxial anisotropy energy can lead to higher coercivity that can enhance the maximum energy product (maximum of the product of $B$ and $H$ in the second quadrant of the $B-H$ curve) and hence the stored magnetostatic energy, which is a standard measure for the performance of a permanent magnet \cite{SKOMSKI2016}. Moreover, MAE can become a key player in stabilizing “skyrmions" in chiral magnets \cite{Bala2020}.

Though synthesized long back \cite{Busey1962, Smith1966, Armstrong1978, Mintz1979}, vacancy-ordered antifluorite compounds with chemical formulae K$_2$BX$_6$, X being a halogen and B being a $5d^{3-5}$ transition metal element, are gaining renewed attention as quantum materials in recent times \cite{Bhaskaran2021,Khan2019,Lee2022,Reig2020}. The Ir-based $5d^5$ antifluorites, K$_2$IrCl$_6$, K$_2$IrBr$_6$, and (NH$_4$)$_2$IrCl$_6$ are intensely studied experimentally and many intriguing properties like antiferromagnetic spin resonance, weak Kitaev anisotropy, non-local cubic distortions, and enhancement of magnetic couplings with halogen replacement are reported \cite{Busey1962,Khan2019,Khan2021,Bhaskaran2021,Lee2022,Reig2020}. Earlier neutron diffraction studies on K$_2$ReCl$_6$ (5$d^3$) have revealed Type-I antiferromagnetic ordering with easy axis along [001] direction. In the case of Os$^{4+}$ ($5d^4$) halides, K$_2$OsCl$_6$, K$_2$OsBr$_6$, Na$_2$OsBr$_6$, and Na$_2$OsBr$_6$·6H2O, either Kotani-like behavior consistent with a $J_{eff}$ = 0 ground state or a weak magnetic ground state is experimentally proposed \cite{Saura2022}.

As a whole, the $5d^{3-5}$ antifluorite compounds provide a platform to examine the complex interplay between various couplings - SOC, electron correlation, magnetic coupling, crystal field, and local exchange field - so that non-trivial phases can be explored. So far, most of the studies on antifluorites have been experimental, and there is a lack of theoretical analysis of electronic and magnetic structures. Specifically, the role of competing interactions in obtaining the magnetic ground state is yet to be explained. 

In this work, we adopt three compounds of the antifluorite family, namely, K$_2$ReCl$_6$, K$_2$OsCl$_6$, and K$_2$IrCl$_6$, and pursued DFT calculations and toy model study to address the following questions: (i) Do antifluorites possess expected pseudo-spin ground state? (ii) How the electronic and magnetic structure evolves due to valence electron count, crystal field, exchange-splitting, and SOC? (iii) Can a large/giant MAE be realized in these systems due to possible spin anisotropies? (iv) Can the ground state and MAE be modified by tuning the crystal field via external stimuli such as strain?

The electronic structure, obtained using density functional theory (DFT) calculations, and model analysis infers that the breakdown of $J_{eff}$ = 3/2 and $J_{eff}$ = 0 pseudo-spin states in KReC and KOsC is due to the presence of large exchange-splitting ($\Delta_{ex}$) compared to the SOC strength ($\Delta_{ex}$ $>>$ $\lambda$). On the contrary, the KIrC stabilizes in $J_{eff}$ = 1/2 state as it possesses weak $\Delta_{ex}$ ($\Delta_{ex}$ $<<$ $\lambda$) so that the SOC led pseudo-spin states remains unperturbed. The nearest-neighbor magnetic exchange interaction $J_1$ is estimated to be $<$ 1 meV, indicating that antifluorites are weakly coupled magnetic systems. While KOsC and KReC stabilize in a Type-I antiferromagnetic ground state with in- and out-of-plane spin orientations, KIrC stabilizes in a Type-III ground state with spin orientation along $\hat{x}$. The easy axis/plane of magnetization is well understood qualitatively as well as quantitatively by carrying out second-order perturbative analysis of SOC and through the quantification of magnetic anisotropic constants. With tensile strain, while the ground state magnetic structure remains robust for KReC and KOsC, it transforms to Type-II with compressive strain irrespective of B-site ion. The magnetic anisotropy energy is found to be large $\approx$ 0.6-4 meV/TM and dominated by single-ion anisotropy energy. Moreover, with strain, the MAE enhances by one order in magnitude.      

\section{Structural and Computational Details}
\begin{figure}
\centering
\includegraphics[angle=-0.0,origin=c,height=6.5cm,width=8cm]{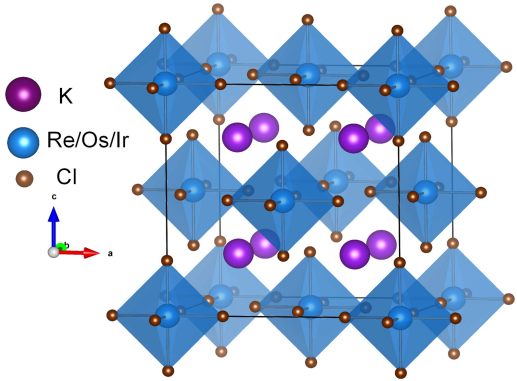}
\caption{The crystal structure of K$_2$BCl$_6$. The structure is realized by introducing vacancies at the B$^\prime$ sites of the double perovskite A$_2$BB$^\prime$O$_6$ crystal structure. Therefore, instead of a corner sharing network of octahedra, in K$_2$BCl$_6$, the fcc stacked BCl$_6$ octahedra are isolated.}
\label{crysal}
\end{figure}

Antifluorites K$_2$ReCl$_6$, K$_2$OsCl$_6$, and K$_2$IrCl$_6$ crystallize in fcc lattice (Fm-3m) (see Fig. \ref{crysal}) at low temperature, with antifluorite-type A$_2$BX$_6$ crystal structure. The crystal structure of antifluorites resembles those of double perovskites (A$_2$BB$^\prime$X$_6$) due to the presence of vacant B$^\prime$ sites. As a result, the BCl$_6$ octahedron is isolated and arranged on a geometrically frustrated fcc lattice as shown in Fig. \ref{crysal}. The optimized lattice parameters are provided in Table-\ref{linear-response}, which are in close agreement with the experimental lattice parameters. 

\begin{figure}
\includegraphics[angle=-0.0,origin=c,height=2.9cm,width=8.7cm]{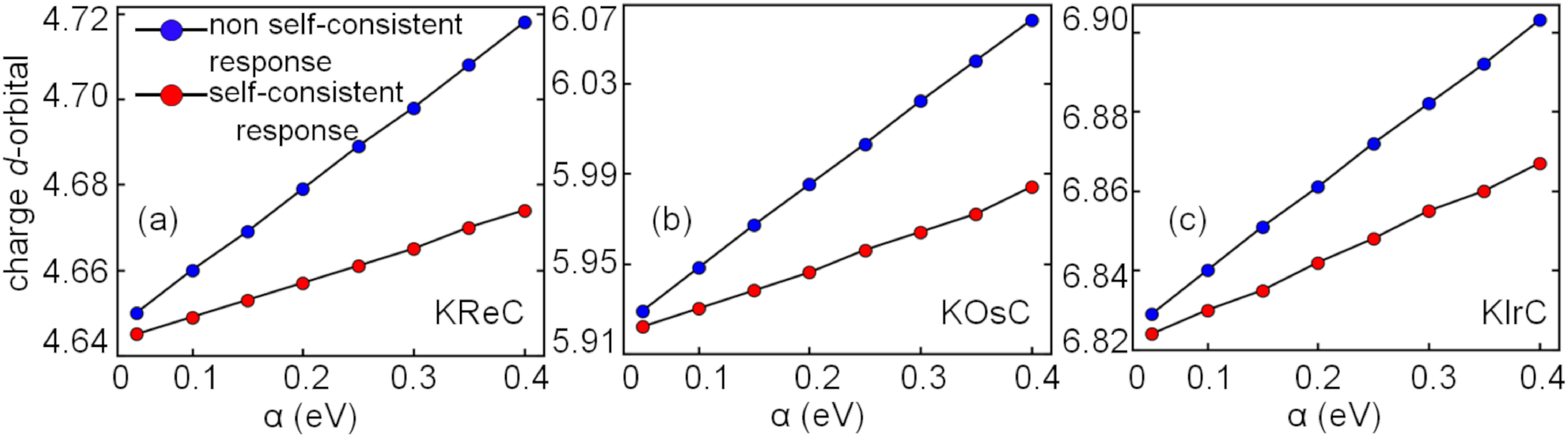}
\caption{(a,b,c) The linear self- and non-self-consistent response functions as a function of the onsite perturbing potential $\alpha$ to estimate the Hubbard $U$ parameter. The solid lines are guide to the eyes.}
\label{U-esti}
\end{figure}

We have adopted DFT+$U$+SOC formalism to study the electronic and magnetic structures. The calculations were performed using plane-wave based projector augmented wave (PAW) \cite{Kresse1999, Bloch1994} method as implemented in Vienna ab-initio simulation package (VASP) \cite{Kresse1996} within the Perdew$-$Burke$-$Ernzerhof generalized gradient approximation (PBE-GGA) for the exchange-correlation functional. The Brillouin zone integrations were carried out using $4 \times 8 \times 8$ $\Gamma$-centered $k$-mesh. The kinetic energy cutoff for the plane-wave basis set was chosen to be $400$ eV. The strong correlation effect was incorporated via an effective onsite correlation parameter $U_\mathrm{eff}$ = $U-J$ through the rotationally invariant approach introduced by Dudarev \cite{Dudarev1998}. For the unstrained case, the calculations were performed on the fully optimized (atomic as well as volume relaxation) experimental structures, whereas, for the strained case, the atomic positions were relaxed while keeping the volume fixed. All optimization calculations were performed by considering onsite correlation $U$ at the TM site.\\
To estimate the Hubbard $U$ parameter, we have employed the linear response approach \cite{Cococcioini2005} wherein the onsite Coulomb repulsion $U$ is computed by calculating the difference between interacting and non-interacting density response functions:
\begin{eqnarray}
    U = \chi^{-1}_0 - \chi^{-1} 
      =  \Big(\frac{\partial n_i^{KS}}{\partial \alpha_i}\Big)^{-1} - \Big(\frac{\partial  n_i}{\partial \alpha_i}\Big)^{-1}.
\end{eqnarray}
The quantity $\alpha$ is the applied perturbation at the TM site where $U$ is computed. The response functions are obtained by performing non-self-consistent ($\chi$) and self-consistent ($\chi_0$) DFT calculations (see Fig. \ref{U-esti}). The obtained values of response functions and Hubbard $U$ are provided in Table-\ref{linear-response}. 

\begin{table}
    \centering
     \caption{The estimated response functions and Hubbard $U$ parameter from linear response theory and experimental, optimized lattice parameters.}
     \vspace{0.2cm}
\begin{tabular}{cccccc} \hline \hline
\vspace{0.2cm}
        Compound & $\chi_0^{-1}$ (eV) & $\chi^{-1}$ (eV) & $U$ (eV) & $a^{exp}$ (\AA) & $a^{relax}$ (\AA)  \\ \hline
        KReC & 12.19 & 5.18 & 7.01 & 9.84 & 9.93\\
        KOsC & 5.74 & 2.71 & 3.03 & 9.68 & 9.78 \\
        KIrC & 8.13 & 4.76 & 3.37 & 9.77 & 9.92 \\
        \hline\hline
    \end{tabular}
    \label{linear-response}
    \end{table}
\section{Electronic and magnetic structure of Antifluorites}
\subsection{Electronic structure}
In this subsection, we will analyze the electronic structure of antifluorites KReC, KOsC, and KIrC wherein the 4+ charge state of TM element leads to $d^3$, $d^4$, and $d^5$ valence states, respectively.\\

\begin{figure}
\includegraphics[angle=-0.0,origin=c,height=17.1cm,width=9cm]{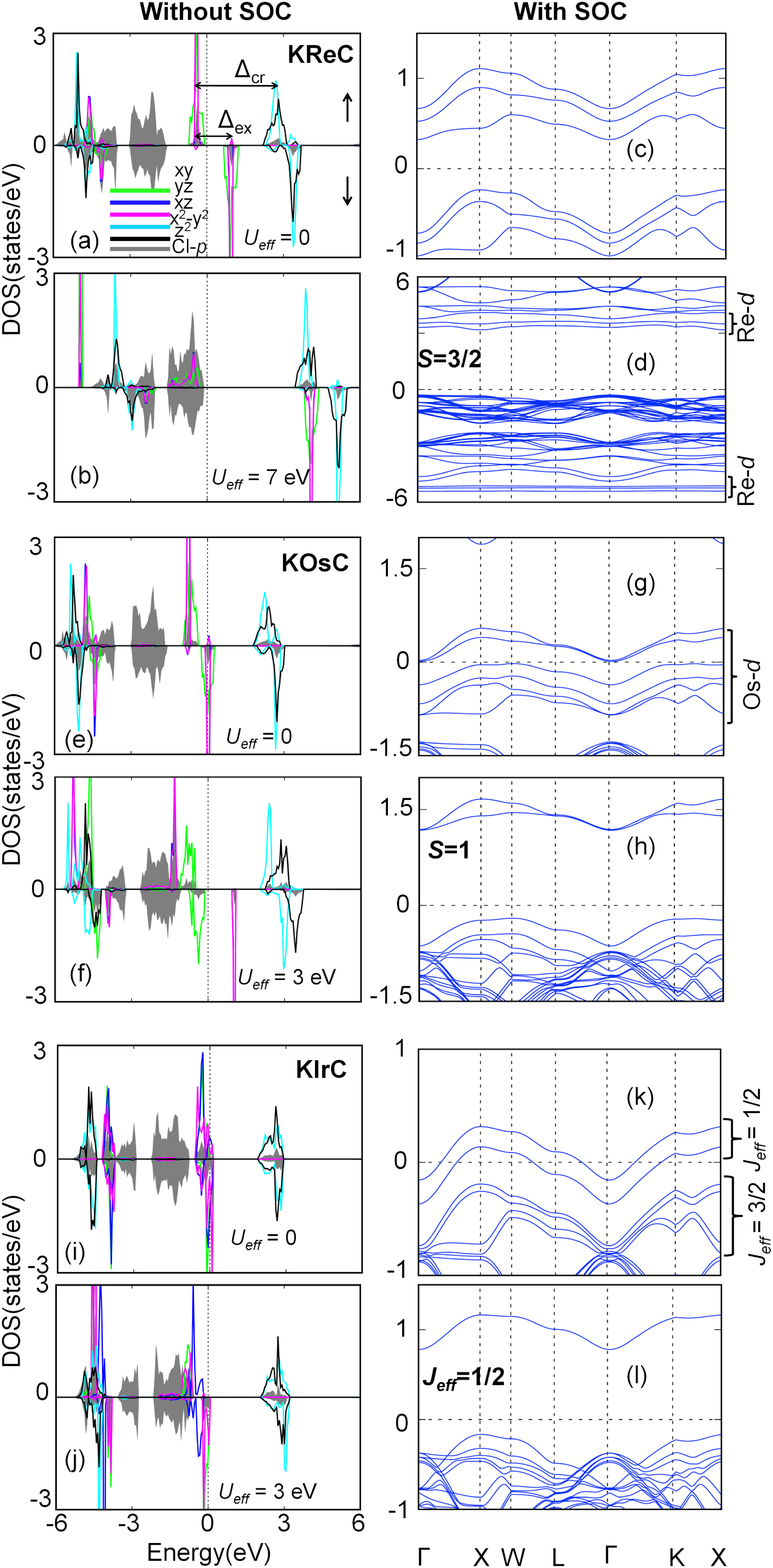}
\caption{Left Column: The spin and orbital resolved density of states per atom with and without considering the effect of onsite Coulomb repulsion in the absence of SOC. Right Column: The corresponding band structures with SOC included. With SOC, the band structures are shown instead of the density of states to elucidate clearly the SOC lead splitting. The $U_{eff}$ values are used as obtained from the linear response method (see Table-\ref{linear-response}).}
\label{ele-struct}
\end{figure}

{KReC-$d^3$:} The spin and orbital resolved partial density of states (pDOS) of KReC is shown in Fig. \ref{ele-struct}(a). The strong localization of the states, even in the absence of onsite repulsion, indicates weak $d-d$ covalent interaction between neighboring TM ions. This is expected because the nearest-neighbor hopping channels are broken due to vacant B$^\prime$ sites, which are otherwise present in the regular double perovskite materials. The effect of pure crystal field splitting on $d$ manifold can be understood by analyzing the spin-polarized GGA pDOS (see Fig. \ref{ele-struct}(a)). 
Due to strong octahedral crystal field ($\Delta_{cr}\approx$ 3 eV estimated by taking out the differences between $t_{2g}$ and $e_g$ band centers), the five-fold degenerate $d$ states split into a higher-energy and unoccupied $e_\mathrm{g}$ doublet ($x^2-y^2$ and $z^2$) and a lower-energy $t_{2g}$ triplet ($xy$, $yz$, and $xz$). While the majority spin-up channel of the latter is completely occupied, the spin-down channel is completely unoccupied, suggesting $t_{2g}^{3\uparrow}$ $e_g^0$ electronic configuration and 4+ charge state of Re. However, the hybridization between Re-$d$ and Cl-$p$ states induces a moment of $\approx$ 0.4 $\mu_B$ on the latter, reducing the effective moment at the Re site (see Table \ref{moments}). The complete separation between the spin-up and -down channels is due to the presence of large exchange-splitting ($\Delta_{ex}$ $\approx$ 1.35 eV), which also led to the formation of an insulating gap even without the inclusion of onsite Coulomb repulsion. Therefore, antifluorite KReC possesses a trivial band-insulating state. With the incorporation of $U_{eff}$ = 7 eV, as obtained from the linear response method (see Table \ref{linear-response}), while the Cl-$p$ states dominate near the Fermi level ($E_F$), the $t_{2g}$ states are pushed deep in the valence and conduction bands, increasing the band gap remarkably to 3.2 eV (see Fig. \ref{ele-struct}(b)).

\begin{table}
\caption{The estimated spin and orbital moments, exchange splitting, and SOC strength with or without incorporating $U$ and SOC.}
\begin{tabular}{lSSSSSSS}
\toprule\toprule
& \multicolumn{2}{c}{GGA} & \multicolumn{3}{c}{GGA+SOC} & \multicolumn{2}{c}{GGA+SOC+$U$} \\
\cmidrule(r){2-3}\cmidrule(l){4-6}\cmidrule(l){7-8}
Compounds & {m$_s$} & {$\Delta_{ex}$} & {m$_s$} & {m$_l$} & {$\lambda$} & {m$_s$} & {m$_l$}\\
\midrule
K$_2$ReCl$_6$ & 2.23 & 1.35 & 2.11 & -0.15 & 0.3 & 2.76 & -0.23 \\
K$_2$OsCl$_6$ & 1.40 & 0.6 & 0.93 & 0.33 & 0.35 & 1.42 & 0.56 \\
K$_2$IrCl$_6$ & 0.35 & 0.1 & 0.20 & 0.34 & 0.43 & 0.28 & 0.48\\
\bottomrule
\end{tabular}
\label{moments}
\end{table}

Having understood the effect of $\Delta_{cr}$ and $U$, we now analyze the effect of SOC on the electronic structure. As shown in Fig. \ref{ele-struct}(c), the SOC lifts the three-fold degeneracy of the $t_{2g}$ manifold to form complex states. As a consequence, the orbital moment (m$_l$) estimated within GGA+SOC turned out to be 0.15 $\mu_B$. Furthermore, with the incorporation of both $U$ and SOC, the Cl-$p$ states form the valence band maxima and dominate near the $E_F$. The m$_l$/m$_s$ ratio is estimated to be 0.07, much less than 1, indicating the quenched effect of SOC. The large spin magnetic moment ($\approx$ 2.8 $\mu_B$) suggests the stabilization of a high spin $S=3/2$ state rather than an ideal $J_{eff}$ = 3/2 state as the ground state of one hole in the $J_{eff}$ = 3/2 manifold can only exhibit a low-spin state. We attribute this breakdown to the presence of a strong local exchange field ($\Delta_{ex} >> \lambda$) which suppresses the atomic SOC strength effectively by $\lambda_{eff}$ = $\lambda_{atomic}/S$ \cite{Khomskii2014}.\\

\begin{figure}
\includegraphics[angle=-0.0,origin=c,height=3.5cm,width=9cm]{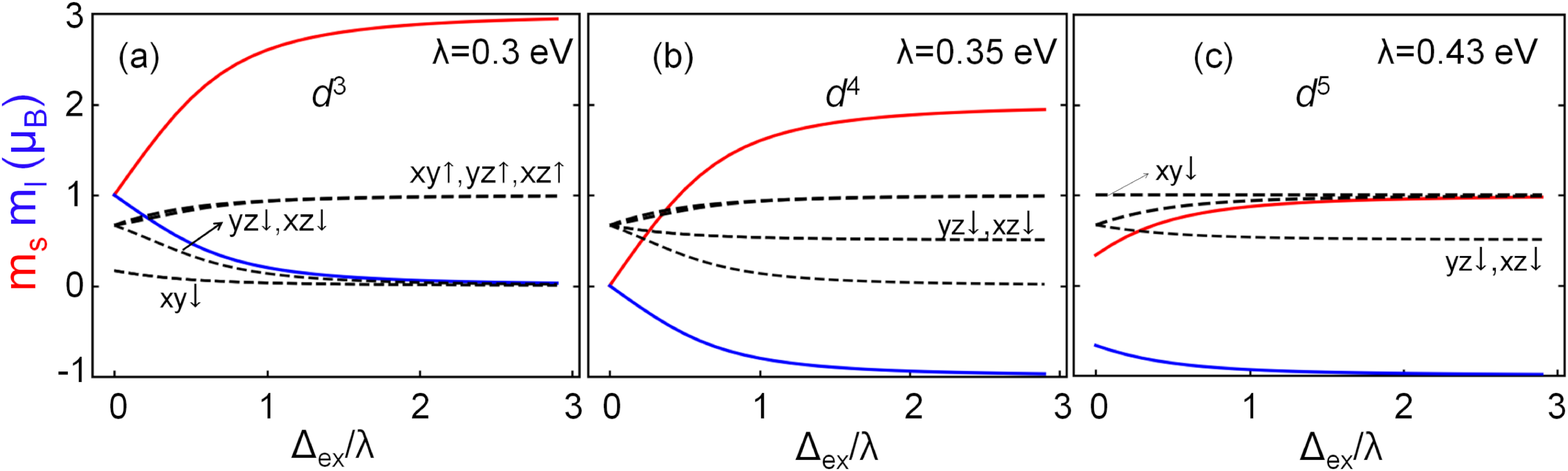}
\caption{The variation in spin (red) and orbital (blue) moments and spin resolved orbital occupation numbers (black dashed lines) as a function of $\Delta_{ex}$/$\lambda$ for $d^3$, $d^4$, and $d^5$ fillings in the atomic limit.}
\label{at-Ham}
\end{figure}

To gain further insights into the breakdown of the $J_{eff}$ = 3/2 state, we now analyze how the pseudo-spin orbitals and the splitting between $J_{eff}$ = 1/2 doublet and $J_{eff}$ = 3/2 quartet evolves as $\lambda$ and $\Delta_{ex}$ compete with each other. In the second quantization formalism, the atomic Hamiltonian is given by,
\begin{equation}
     H^{at}=\lambda \sum_{\alpha,\beta,\sigma,\bar{\sigma}} \bra{\alpha \sigma} \boldsymbol{L} \cdot \boldsymbol{S} \ket{\beta \bar{\sigma}} c_{\alpha, \sigma}^\dagger c_{\beta,\bar{\sigma}}\\
      +\Delta_{ex}\sigma\sum_{i,\alpha,\sigma}c^\dagger_{i\alpha\sigma}c_{i\alpha\sigma},
\end{equation}
where $\alpha$ ($\beta$) and $\sigma$ represent orbitals ($xy$, $yz$, and $xz$) and spin ($\uparrow,\downarrow$) indices, respectively. The obtained results are shown in Figs. \ref{orbitals} (c) and \ref{at-Ham} (a) wherein the energy, spin/orbital moments, and occupation numbers are plotted as a function of $\Delta_{ex}$/$\lambda$. 
While in $\lambda$$>>$$\Delta_{ex}$ limit, SOC entangles the $t_{2g}$ states with respect to spin and orbital to form pseudo-spin states (see Fig. \ref{orbitals} (b)), 
\begin{align}
  \ket{\frac{1}{2}, \pm{\frac{1}{2}}} &= \frac{1}{\sqrt{3}}( \ket{yz,\bar{\sigma}} \pm{\ket{xy,\sigma}} \pm{i} \ket{xz,\bar{\sigma}})\\
   \ket{\frac{3}{2}, \pm{\frac{1}{2}}} &= \frac{1}{\sqrt{6}}(\ket{yz,\bar{\sigma}} \mp 2{\ket{xy,\sigma} } \pm{i} \ket{xz,\bar{\sigma}})\\
    \ket{\frac{3}{2}, \pm{\frac{3}{2}}} &= \frac{1}{\sqrt{2}}(\ket{yz,\sigma} \pm{i} \ket{xz,{\sigma}})
  \end{align}
, the increasing strength of $\Delta_{ex}$/$\lambda$ breaks the four-fold and two-fold degeneracy of $J_{eff}=3/2$ and $J_{eff}=1/2$ states and leads to complete decoupling of the spin-up and -down states in $\lambda$$<<$$\Delta_{ex}$ limit (see Fig. \ref{orbitals}(b)). The decoupling between the former and the latter occurs as the occupation numbers of the $t_{2g}^\uparrow$ orbitals increase with $\Delta_{ex}$/$\lambda$ whereas the occupation numbers of the 
$t_{2g}^\downarrow$ orbitals decreases and becomes zero (see Fig. \ref{at-Ham} (a)). This leads to the complete unquenching and quenching of spin and orbital moments ($m_s$ = 3 $\mu_B$ and $m_l$ = 0). Therefore, $d^3$ configuration, which is expected to possess large $\Delta_{ex}$, is bound to stabilize in a high-spin $S=3/2$ state rather than weakly spin-polarized $J_{eff}$ = 3/2 state. To understand the vanishing orbital moment with $\Delta_{ex}$, we analyze the eigen states of $H^{at}$ in $\Delta_{ex}$ $>>$ $\lambda$ limit. As inferred from Fig. \ref{orbitals} (b), while $\ket{3/2,\pm 3/2}$ orbitals remains unperturbed by $\Delta_{ex}$, the $\ket{3/2,\pm 1/2}$ and $\ket{1/2,\pm 1/2}$ orbitals evolve to $xy\uparrow$, $yz\uparrow - ixz\uparrow$, $xy\downarrow$, and $yz\downarrow + ixz\downarrow$ orbitals, respectively. Since, $\ket{3/2,3/2}$ and $yz\uparrow-ixz\uparrow$ orbitals possess equal and opposite orbital polarization ($\braket{3/2,3/2|\hat{L_z}|3/2,3/2}$ = 2, $\braket{yz\uparrow - i xz\uparrow|\hat{L_z}|yz\uparrow - ixz\uparrow}$ = -2, and $\braket{xy|\hat{L_z}|xy}$ = 0), the net orbital moment becomes zero. However, the spin polarization of lower-lying three orbitals is parallel, as a result, $S$ = 3/2 state forms.

Invoking an intermediate coupling approach that incorporates both SOC and $U$ on an even footing, RIXS measurements on Ca$_3$LiOsO$_6$ and Ba$_2$YOsO$_6$ (Os-$d^3$) systems report $J_{eff}$ = 3/2 ground state \cite{Taylor2017}. However, first-principles studies on these systems found $U$, $\Delta_{ex}$ $>>$ $\lambda$. Therefore, as established from our analysis for $d^3$ configuration in $\Delta_{ex}$ $>>$ $\lambda$ limit, Ca$_3$LiOsO$_6$ and Ba$_2$YOsO$_6$ will exhibit high-spin $S$ = 3/2 ground state rather than $J_{eff}$ = 3/2 pseudo-spin state. 

KOsC-$d^4$: The $d^4$ configuration is expected to give rise to a nonmagnetic $J_{eff}$ = 0 insulating state due to completely occupied $J_{eff}$ = 3/2 manifold (see Fig. \ref{orbitals} (a)). However, contrary to the expected ground state, many candidate materials were found to be stabilized in a broken $J_{eff}$ = 0 state \cite{Bandyopadhyay2022, Bhowal2015}. To find whether it breaks down or stabilizes in KOsC, we examine its electronic structure, which is shown in Figs. \ref{ele-struct} (e-h). For the spin-polarized GGA case, the $t_{2g}$ states are completely occupied in the spin-up channel, whereas, in the spin-dn channel, it is one-third occupied, leading to a half-metallic state.
\begin{figure*}
\includegraphics[angle=-0.0,origin=c,height=4cm,width=15.3cm]{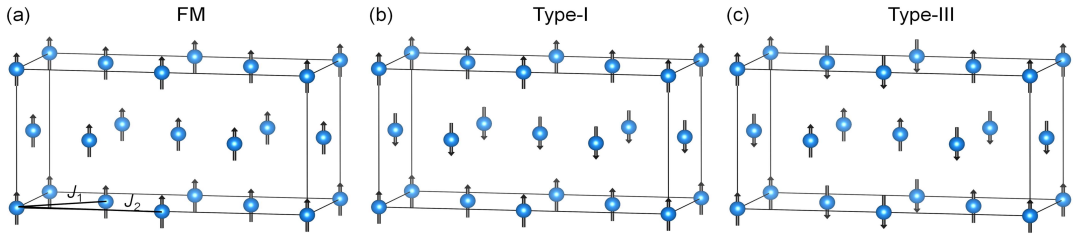}
\caption{(a) The magnetic orderings, namely, FM, Type-I, and Type-III, considered for the estimation of magnetic exchange interactions $J_1$ and $J_2$.}
\label{mag-ord}
\end{figure*}
The half-metallic scenario for $U_{eff}$ = 0 occurs due to weak $\Delta_{ex}$ $\approx$ 0.6 eV. The partial occupancy at the $E_F$ gets lifted with the onset of Coulomb repulsion as it separates the $xy$ and $yz$, $xz$ states with a gap in between (see Fig. \ref{ele-struct}(f)). Therefore, antifluorite KOsC exhibits a Mott insulating state with an intermediate spin state $t_{2g}^{3\uparrow 1\downarrow}e_g^0$. While the gap in KOsC is mediated by strong onsite Coulomb repulsion, SOC led splitting of states at the $E_F$ leads to a semimetallic state. Within GGA+SOC+$U$, while the orbital moment enhances from 0.33 to 0.56 $\mu_B$, the spin moment gets quenched by 0.4 $\mu_B$, giving rise to m$_l$/m$_s$ ratio of $\approx$ 0.36. The m$_l$/m$_s$ ratio increases as compared to KReC as the $\lambda_{eff}$ increases with a decrease in the spin moment. The estimated large spin moment ($\approx$ 1.42 $\mu_B$) indicates the breakdown of the expected $J_{eff}$ = 0 nonmagnetic state in KOsC. Our results are contrary to a recent experimental study which report a $J_{eff}$ = 0 or a weakly magnetic ground state for K$_2$OsCl$_6$ \cite{Saura2022}.

In the atomic limit, as shown in Fig. \ref{at-Ham}(b), for $\Delta_{ex}$ = 0, the spin and orbital moments are zero and as a result a $J_{eff}$ = 0 state forms. This occurs as the $J_{eff}$ = 3/2 states possess equal and opposite spin and orbital polarization for $m_J=\pm1/2$ or $m_J=\pm3/2$ states which are completely occupied. With increasing $\Delta_{ex}$/$\lambda$, the spin and orbital moment increases and saturate at 2 and 1 $\mu_B$, respectively. The former and the latter occurs as yz$\downarrow$ and xz$\downarrow$ orbitals are half-occupied to form $yz\downarrow$ - ixz$\downarrow$ orbital which possess $m_s$ = -1, $m_l$ = -1 moments while the rest of the three lower-lying orbitals gives $m_s$ = 3 and $m_l$ = 0 (see Fig. \ref{orbitals} (b)). As inferred from Table-\ref{moments}, KOsC possess $\Delta_{ex}$/$\lambda$ ratio of $\approx$ 1.71. As a consequence, $S$ = 1 state stabilizes instead of the nonmagnetic $J_{eff}$ = 0  state. Therefore, we attribute the breakdown of the $J_{eff}$ = 0 state in KOsC to the presence of substantial local exchange field.\\    

KIrC-$d^5$: The pDOS for KIrC are shown in Figs. \ref{ele-struct} (i-j). For the GGA case, both spin channels are more or less equally occupied and hence a low-spin state resulting from weak $\Delta_{ex}$ $\approx$ 0.1 eV stabilizes. With the inclusion of $U$, while the the partially occupied states in the spin-up channel are now completely occupied, the spin-dn channel remains partially occupied. Therefore, unlike the case of KOsC, the inclusion of $U$ does not lead to opening of a gap at the $E_F$. With SOC, the $t_{2g}$ states spin mixes to form completely occupied $J_{eff}$ = 3/2 states and partially occupied $J_{eff}$ = 1/2 states (see Fig. \ref{ele-struct} (k)). Interestingly, a large gap opens up within GGA+SOC+$U$, as the combined effect of correlation and SOC lifts the partially occupancy of the $J_{eff}$ = 1/2 states to open up a gap. Therefore, antifluorite KIrC possesses a non-trivial spin-orbit-assisted Mott insulating state which is yet not reported in the literature. The opening of a gap manifests the pronounced effect of SOC as the weak Ir spin moment $\approx$ 0.28 $\mu_B$ enhances the effective SOC with m$_l$/m$_s$ ratio of $\approx$ 1.71.\\

As shown in Fig. \ref{orbitals} (b), the $J_{eff}$ = 1/2 ($m_j$ = 1/2) orbital transform to $xy\downarrow$ orbital in large $\Delta_{ex}$/$\lambda$ limit. Since $xy$ orbital possess $m_l$ = 0, the addition of spin and orbital moments of five lower-lying states give rise to $m_s$ = 1 and $m_l$ = -1. However, the former is the least probable scenario as one hole in the $t_{2g}$ manifold tends to stabilize the system in a low-spin state. As $\Delta_{ex}$/$\lambda$ in KIrC is very weak ($\approx$ 0.23), the SOC-led pseudo-spin states remain more or less unperturbed to form a weakly spin-polarized $J_{eff}$ = 1/2 ground state.

\subsection{Magnetic structure of Antifluorites and its strain engineering}
We now proceed to investigate the ground state magnetic structure of antifluorites and its tuning with strain. Generally, depending on the strength and nature of the nearest and next-nearest-neighbor spin exchange interactions ($J_1$ and $J_2$) \cite{Henley1987,Singh2017,Lines1963}, the spin arrangement on a geometrically frustrated fcc lattice can be of Type-I, Type-II or Type-III (see Fig. \ref{mag-ord})\cite{Henley1987,Singh2017,Lines1963}.
\begin{figure*}
\includegraphics[angle=-0.0,origin=c,height=7.2cm,width=13.4cm]{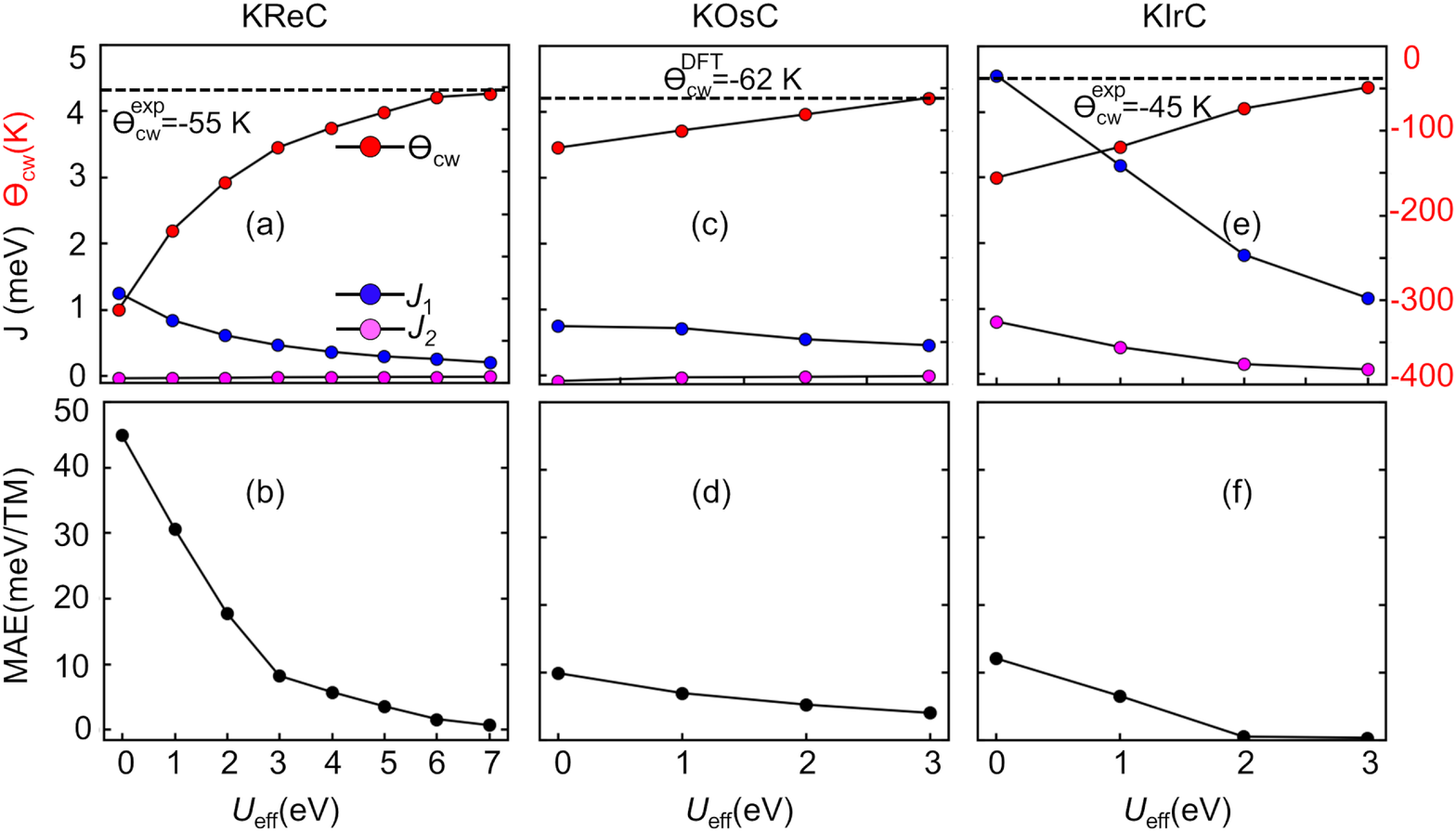}
\caption{Upper panel: The variation of magnetic exchange parameters $J_1$, $J_2$, and Curie Weiss temperature $\theta_{cw}$ as a function of $U_{eff}$. Lower panel: The magnetic anisotropy energy/transition metal atom as a function of $U_{eff}$. The $U_{eff}$ is varied upto the theoretically calculated value from the linear response theory (see Table \ref{linear-response}).}
\label{bulk-J}
\end{figure*}
Therefore, in order to examine the magnetic structure of antifluorites, we estimate the exchange interactions by employing Noodelmann’s broken-symmetry spin dimer method \cite{Noodleman1981}. In this method, the energy difference between high spin (HS) and broken symmetry (BS) configurations is given by
\begin{equation}
E_{HS} - E_{BS} = \frac{1}{2} S_{max}^2 J
\end{equation}
where $J$ is related to the spin-dimer Hamiltonian $H$ = $\sum_{i<j} J_{ij} S_i \cdot S_j$ and $S_{max}$ is the maximum spin of the dimer. For the present case, since each monomer has three (KReC-$d^3$), two (KOsC-$d^4$), and one (KIrC-$d^5$) unpaired spins, the corresponding $S_{max}$ values are 3, 2, and 1, respectively. The $E_{HS}$ and $E_{BS}$ are evaluated by performing the DFT calculations for HS and BS configurations. To estimate $J$, we have adopted three different spin configurations: FM, Type-I, and Type-III (see Fig. \ref{mag-ord}). We have not considered Type-II magnetic ordering because the minimum supercell size required to design it is 2$\times$2$\times$2 which contains four formulae units of K$_2$BCl$_6$, thereby making the estimation of $J$ computationally expensive. By employing the spin dimer Hamiltonian on these magnetic configurations, we have calculated the spin-exchange energies (per f.u.) in terms of the spin-exchange parameters, which are expressed as \par
\begin{eqnarray}
E_{FM} = \frac{S^2}{2}(48J_1 + 40J_2),\notag\\
E_{I} = \frac{S^2}{2}(-16J_1 + 40J_2),\notag\\
E_{III} = \frac{S^2}{2}(-16J_1 + 24J_2 ),
\end{eqnarray}
where $S$ is the spin of each monomer. The spin-exchange energies are mapped to the total energy obtained from the DFT calculations for the respective magnetic configuration. Further, using Eqns. 6-7, the $J$ values and Curie-Weiss temperature ($\Theta_{CW} = - \frac{S(S+1)}{3K_B} \sum_i Z_i J_i$ where $Z_i$ is coordination number corresponding to each $J_i$
interaction) are estimated as a function of $U$ and the results are shown in the top row of Fig. \ref{bulk-J}. 

As inferred from Fig. \ref{bulk-J}, the $\Theta_{CW}$ is negative, indicating that the dominant interaction is antiferromagnetic. The $\Theta_{CW}$ or the Neel temperature $T_N$ are often evaluated as a function of $U$ and matched with the experimental results. From this matching, a realistic value of $U$ can be determined to re-validate the $U$ strength obtained from the linear response method. The experimentally reported values of $\Theta_{CW}$ for KReC \cite{Busey1962} and KIrC \cite{Khan2019} are -55 K and -45 K, respectively, which matches close to the theoretically estimated values $\approx$ -58.15 K and -49.05 K at $U$ = 7 and 3 eV, respectively. These values are in excellent agreement with those obtained from the linear response method (see Table \ref{linear-response}). For KOsC, to the best of our knowledge, the experimental value of $\Theta_{CW}$ is not available in the literature. Therefore, its $\Theta_{CW}$ value can be roughly approximated to -62 K at $U_{eff}$ = 3 eV.

The sharp decrease in $J_1$ with increasing $U_{eff}$ manifests the superexchange interaction between NN spin moments. For KReC, at $U_{eff}$ = 7 eV, the estimated strengths of $J_1$ and $J_2$ ($\approx$ 0.2 and -0.02 meV) suggest that the NN(NNN) magnetic interactions are weak and antiferromagnetic(ferromagnetic) in nature. The DFT estimated $J$ strengths are in good agreement with those estimated using correlated effective-field theory \cite{Chatterjee1981}. The obtained nature of exchange interactions leads to the stabilization of a Type-I magnetic order \cite{Henley1987} consistent with the neutron diffraction studies \cite{MINKIEWICZ1968}.
\begin{table}
    \centering
     \caption{The estimated energy differences (meV/TM) with respect to the most stable magnetization direction within DFT+$U$+SOC calculations. The single ion anisotropy energies are given in parentheses.}
     \vspace{0.2cm}
\begin{tabular}{cccc} \hline \hline
        SAXIS & KReC & KOsC & KIrC\\ \hline
        $[001]$ & 0 & 3.97 (3.8) & 0.18 \\
        $[100]$ & 0.20 & 0.44 & 0 \\
        $[110]$ & 0.58 & 0 & 0.49\\
        $[111]$ & 0.64 (0.51) & 0.27 & 0.47 (0.28)\\
        \hline\hline
    \end{tabular}
    \label{aniso-energy}
    \end{table}
To determine the easy/hard axis of magnetization, we perform DFT calculations with the spin quantization axis (SAXIS) pointing along different directions. The obtained energy differences are listed in Table-\ref{aniso-energy}. As inferred from Table \ref{aniso-energy}, the easy and hard axis for KReC is along [001] and [111] directions, differing in energy, the so-called magnetic anisotropy energy, by $\approx$ 0.64 meV/Re. The underlying origin behind the preferred spin orientation will be discussed in the next section. Like KReC, KOsC also exhibits Type-I antiferromagnetic order with weak $J_1$ and $J_2$ ($\approx$ 0.45 and -0.02 meV). However, KOsC possesses an easy plane [110] of magnetization which is stabilized by large MAE of $\approx$ 3.97 meV/Os. For KIC, the $J_1$ ($\approx$ 1.2 meV) and $J_2$ ($\approx$ 0.07 meV) are both antiferromagnetic and hence KIC exhibits a Type-III order \cite{Henley1987} with spin orientation along $\hat{x}$. Moreover, the ground state remains robust with increasing $U_{eff}$ as the nature of interactions remains the same.

Often strain is used as an external stimuli to tune the magnetic properties and transition temperatures. To observe the effect of it on the ground state of antifluorites, we have applied an epitaxial tensile and compressive strain and the results are shown in Fig. \ref{strain}. With tensile strain, while the nature of $J_1$ and $J_2$ and hence the magnetic order remains robust for KReC and KOsC, the easy/hard axis for the former and latter switches to [110]/[001 and [100]/[001], respectively (see Figs. \ref{strain}(b-d)). However, for KIrC, spin ordering transforms from Type-III to Type-I ($J_2$$<$0) with easy/hard axis along [001] and [110] directions. For compressive strain, the $J_2$ enhances significantly as compared to $J_1$, therefore, a Type-II order \cite{Henley1987} stabilizes irrespective of the B-site element. Moreover, while the easy and hard axis remains robust for KIrC, it changes to in- and out-of-plane for KReC and KOsC, respectively. 

\section{MAE and preferred spin orientation}
Magnetic anisotropy energy is an important quantity and can lead to higher coercivity that can enhance the performance of permanent magnets \cite{SKOMSKI2016}. It is estimated by taking the energy difference between easy and hard directions (MAE = $E^{easy}$ - $E^{hard}$) of magnetization for a magnetic material. Since the antifluorites considered in this study involve heavier TM and hence possess substantial SOC, which is a key player to realize large/giant MAE, it is worth examining MAE in these compounds. Taking the easy axis as the reference direction, the calculated values of MAE are listed in Table \ref{aniso-energy}. The MAE is large and is at least one-to-two orders higher than traditional MAE materials like transition metals
and their multilayers \cite{Chauhan2021}. 
\begin{figure*}
\includegraphics[angle=-0.0,origin=c,height=7.2cm,width=13cm]{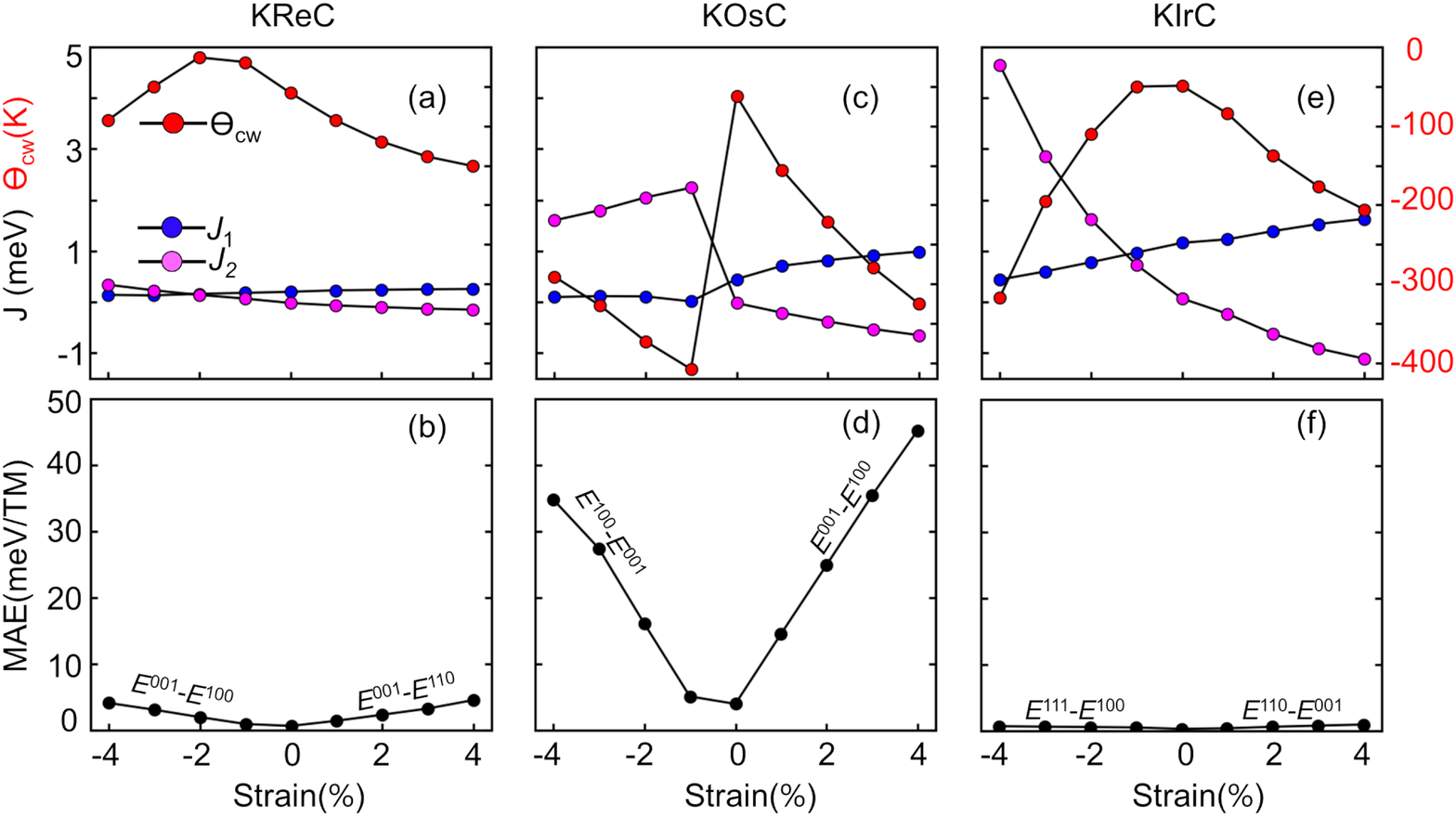}
\caption{Upper panel: The variation of magnetic exchange parameters $J_1$, $J_2$, and Curie Weiss temperature $\theta_{cw}$ as a function of epitaxial tensile and compressive strain. Lower Panel: The magnetic anisotropy energy as a function of strain. The $U_{eff}$ is kept fixed as obtained from linear response theory.}
\label{strain}
\end{figure*}

While we found antifluorites to be promising candidates for the realization of large/giant MAE, we now examine the SOC-induced interaction that contributes dominantly to MAE. The SOC-induced interactions include antisymmetric exchange or Dzyaloshinskii-Moriya (DM) interaction, anisotropic exchange, and single-ion anisotropy (SIA). Due to the presence of inversion symmetry, the DM interaction will not contribute to MAE. As the antifluorites possess weak isotropic $J$, the contribution to MAE from the anisotropic exchange will also be negligible. Therefore, presumably, SIA will contribute dominantly to the estimated MAE values. To substantiate it further, we estimate SIA by replacing the neighboring Ir, Os, and, Re atoms in KReC, KOsC, and KIrC with non-magnetic Si ions. In this way, the contribution from $J_1$ and $J_2$ interactions will be killed in the total energy calculations. The obtained SIA values are provided in the bracketed terms of Table \ref{aniso-energy}. As expected, we found the SIA to be the dominant contributor to MAE.   

Further, to study the strain tunability of MAE, we apply epitaxial strain, and the results are shown in the lower panel of Fig. \ref{strain}. The MAE increases linearly with strain and becomes one-order higher with respect to the ground state value. Intriguingly, for Os, the MAE becomes giant ($\approx$ 20-40 meV/Os for 2-4 $\%$ strain).

The preferred ground state spin orientation (in or out-of-plane) of antifluorites can be analyzed by employing second-order perturbative analysis of SOC Hamiltonian $H_{SO}$ = $\lambda$ $\hat{L}$ $\cdot$ $\hat{S}$. Defining two independent coordinate systems (x, y, z) and (x$^\prime$, y$^\prime$, z$^\prime$) for the orbital and spin angular momentum $\hat{L}$ and $\hat{S}$, $H_{SOC}$ can be rewritten as $H_{SOC}$ = $H_{SOC}^{sc}$ + $H_{SOC}^{snc}$, where the spin-conserving ($H_{SOC}^{sc}$) and spin-non-conserving ($H_{SOC}^{snc}$) terms can be written as \cite{Dai2008,Chauhan2021,Wang1993,Wang1996}
\begin{eqnarray}
    H_{SOC}^{sc} &= \lambda S_{z^\prime}(L_z \cos{\theta}+\frac{1}{2}L_{+}e^{-\iota\phi}\sin{\theta}+\frac{1}{2}L_{-}e{^{\iota\phi}}\sin{\theta})\notag\\
    &= \lambda S_{z^\prime}(L_z \cos{\theta}+L_{x}\sin{\theta}\cos{\phi}+L_{y}\sin{\theta}\sin{\phi})\notag \\
\end{eqnarray}
\begin{eqnarray}
    H_{SOC}^{snc} = \frac{\lambda}{2} S_{+}^\prime(-L_z\sin{\theta}-L_{+}e^{-\iota\phi}\sin^2{\theta/2}+L_{-}e^{\iota\phi}\cos^2{\theta/2})\notag \\
    +\frac{\lambda}{2} S_{-}^\prime(-L_z\sin{\theta}+L_{+}e^{-\iota\phi}\cos^2{\theta/2}-L_{-}e^{\iota\phi}\sin^2{\theta/2})\notag \\
    =\frac{\lambda}{2}(S_{+}^\prime+S_{-}^\prime)(-L_z\sin{\theta}+L_x\cos{\theta}\cos{\phi}+L_y\cos{\theta}\sin{\phi})\notag
    \end{eqnarray}
where $\theta$ and $\phi$ determine the magnetization direction with respect to (x,y,z) coordinate system. It is important to note that while $H_{SOC}^{sc}$ gives non-zero expectation values for the same spin states due to $\hat{S_z}$, $H_{SOC}^{nsc}$ gives finite contribution for opposite spin states.

Using perturbation theory, the second-order correction in energy due to SOC is given by,
\begin{equation}
    \Delta E_{soc}=-\lambda^2\sum_{o^\sigma,u^{\sigma^\prime}}\frac{|\bra{o^{\sigma}}H_{soc}\ket{u^{\sigma^\prime}}|^2} {|\epsilon_{o^{\sigma}} - \epsilon_{u^{\sigma^\prime}}|}
\end{equation}
where $u^\sigma$ ($o^{\sigma'}$) denotes the unoccupied (occupied) states in spin ($\sigma$, $\sigma^\prime$  = {$\uparrow$, $\downarrow$}) channels. A qualitative analysis of the spin orientation can be obtained by analyzing the energy separation and coupling to the SOC Hamiltonian between $o^{\sigma}$ and $u^\sigma$ states. The energy difference $\epsilon_{u^\sigma}$ - $\epsilon_{o^{\sigma'}}$ is roughly estimated through the spin and orbital resolved density of states shown in the first column of Fig. \ref{ele-struct}. 

For KReC, the occupied ($t_{2g}\uparrow$) and unoccupied ($t_{2g}\downarrow$+$e_{g}\uparrow\downarrow$) states contributing to MAE are $xy\uparrow$, $yz\uparrow$, and $xz\uparrow$, and $xy\downarrow$, $yz\downarrow$, $xz\downarrow$, $x^2-y^2\uparrow\downarrow$, and $z^2\uparrow\downarrow$, respectively. Since the coupling between $t_{2g}$ and $e_{g}$ states occur in both spin channels ($t_{2g}\uparrow$-$e_{g}\uparrow\downarrow$), the net contribution to MAE turned out to be zero. The $t_{2g}$ states couple in opposite spin-channel, hence, the contribution from $t_{2g}\uparrow - t_{2g}\downarrow$ coupling to MAE comes from the spin-non-conserving term of $H_{SO}$. Using non-vanishing matrix elements $\bra{xz}\hat{L_z}\ket{yz} = 1$, $\bra{xy}\hat{L_x}\ket{xz} = 1$ and Eq. 9, MAE and is given by,
\begin{eqnarray}
    MAE^{[001]-[111]}_{KReC} =\frac{\lambda^2}{4} \sum_{o^\sigma,u^{\sigma^\prime}}\big(
    \frac{|\bra{o^{\sigma}}\sqrt{2}\hat{L_x}+\sqrt{2}\hat{L_y}-2\hat{L_z}\ket{u^{\sigma^\prime}}|^2} {8 \epsilon_{o^{\sigma}}-\epsilon_{u^{\sigma^\prime}}} \notag \\ 
    -\frac{|\bra{o^{\sigma}}\hat{L_x}\ket{u^{\sigma^\prime}}|^2} {\epsilon_{o^{\sigma}}-\epsilon_{u^{\sigma^\prime}}} \notag \big)\\
     = \frac{\lambda^2}{4}\big(\frac{1}{4\epsilon_{xy\uparrow-yz\downarrow}}-\frac{7}{4\epsilon_{xy\uparrow-xz\downarrow}}\notag \\
    +\frac{1}{4\epsilon_{yz\uparrow-xz\downarrow}}\big) < 0. \notag \\   
\end{eqnarray}
Since the energy differences between the band centers of the occupied and unoccupied $t_{2g}$ orbitals are the same, the MAE of KReC turned out to be negative. Hence, the [001] magnetization direction is energetically favored.\\

For KOsC, while the contribution to MAE from $t_{2g}\uparrow\downarrow$-$e_{g}\uparrow\downarrow$ and xy$\uparrow$$\downarrow$-yz/xz$\downarrow$ couplings is negligible, the finite contribution comes from yz/xz $\uparrow$-yz/xz$\downarrow$ interactions and is given by,
\begin{eqnarray}
    MAE^{[001]-[110]}_{KOsC} =\frac{\lambda^2}{4} \sum_{o^\sigma,u^{\sigma^\prime}}
    \frac{|\bra{o^{\sigma}}L_z-L_x\ket{u^{\sigma^\prime}}|^2} {     \epsilon_{o^{\sigma}}-\epsilon_{u^{\sigma^\prime}}} \notag \\
    = \frac{\lambda^2}{2\epsilon_{yz\uparrow-xz\downarrow}} > 0. \notag \\
\end{eqnarray}
Therefore, KOsC possesses an easy plane of magnetization. 

Similarly, for KIrC, the MAE is given by,
\begin{eqnarray}
   MAE^{[100]-[111]}_{KIrC} = \frac{\lambda^2}{4}(\frac{1}{2\epsilon_{yz\downarrow-xz\downarrow}}+\frac{1}{4\epsilon_{yz\downarrow-xy\downarrow}}\notag \\
    -\frac{3}{4\epsilon_{xz\downarrow-xy\downarrow}}\notag
     ) < 0,\notag\\
\end{eqnarray}
and as a result, the easy axis lies along $\hat{x}$. 

Qualitatively, the origin of large MAE in KOsC compared to KReC and KIrC can be understood from the evaluated MAE expressions. For KReC and KOsC, while terms contributing positively as well as negatively to MAE are present to lower the MAE magnitude, for KOsC, only positively contributing term is present. Hence, KOsC exhibits a large MAE.  \\
\begin{table}
    \centering
     \caption{The estimated anisotropic constants to determine the easy axis/plane of magnetization of antifluorites.}
     \vspace{0.2cm}
\begin{tabular}{ccc} \hline \hline
        Compound & $K_1$ (meV) & $K_2$ (meV)\\ \hline
         KReC & 1.68 & -1.47\\
         KOsC & -7.32 & 3.72\\
         KIrC & 1.41 & -1.61\\
        \hline\hline
    \end{tabular}
    \label{anisotropic constants}
    \end{table}   
A further quantitative approach to identify the easy axis/plane is through the estimation of anisotropic constants $K_1$ and $K_2$ involved in the expression of crystal anisotropy energy $E(\theta$) = $K_1$ Sin$^2$ $\theta$ + $K_2$ Sin$^4$ $\theta$.
\begin{figure}
\centering
\includegraphics[angle=-0.0,origin=c,height=5cm,width=6cm]{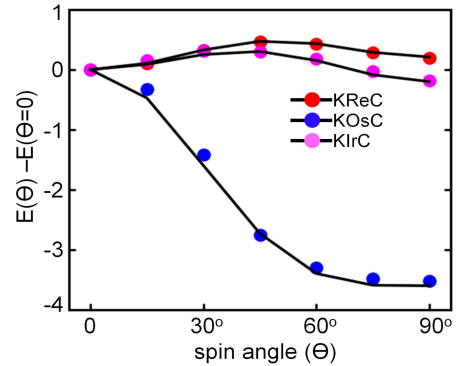}
\caption{The variation in energy difference between spin direction along $\phi$=0 and $\hat{z}$  as a function of the polar angle $\theta$.}
\label{K-ani}
\end{figure}
These constants are estimated by fitting the energy differences $E(\theta)$-$E(\theta=0)$ obtained by rotating the magnetization axis from [001] to [100] directions from DFT calculations (the blue, red, and magenta-filled circles in Fig. \ref{K-ani}) with the parametric equation (black solid line). The obtained values of $K_1$ and $K_2$ are listed in Table-\ref{anisotropic constants}. The conditions for easy $c$-axis (basal-plane) are $K_2$ $>$ -$K_1$ ($K_2$ $<$ -$K_1/2$) \cite{Book2}, which are very well satisfied by KReC (KOsC and KIrC). Therefore, the second-order perturbative analysis of SOC and quantification of anisotropic constants provide good insights into the magnetization direction and large/giant MAE obtained from total energy DFT calculations.\\

\section{Summary}
To summarize, by pursuing density functional theory calculations and a toy model, we examine the electronic and magnetic structure of recently emerging vacancy-ordered antifluorites K$_2$ReCl$_6$ (Re-$d^3$), K$_2$OsCl$_6$ (Os-$d^4$), and K$_2$IrCl$_6$ (Ir-$d^5$). The competition between spin-orbit coupling $\lambda$ and local exchange field $\Delta_{ex}$ suggests that the breakdown of $J_{eff}$ = 3/2 and $J_{eff}$ = 0 states in KReC and KOsC is driven by the presence of large $\Delta_{ex}$ ($\Delta_{ex}$ $>>$ $\lambda$) respectively. As a consequence, while KReC exhibits high-spin $S$ = 3/2 trivial band insulating state, KOsC stabilizes in an intermediate-spin $S$ = 1 Mott insulating state. On the other side, due to weak $\Delta_{ex}$ ($\Delta_{ex}$ $<<$ $\lambda$), the $J_{eff}$ = 3/2 and $J_{eff}$ = 1/2 manifolds remains unperturbed to stabilize KIrC in $J_{eff}$ = 1/2 spin-orbit-assisted Mott insulating state. The nearest and next-nearest-neighbor (NN and NNN) spin-exchange interactions ($J_1$ and $J_2$) estimated using the spin-dimer method infers that the antifluorites are very weakly coupled magnetic systems. For KReC and KOsC, the $J_1$ and $J_2$ are found to be antiferromagnetic and ferromagnetic, leading to a Type-I magnetic ground state, whereas, for KIrC, both $J_1$ and $J_2$ are antiferromagnetic and hence a Type-III ground state stabilizes in KIrC. The estimated anisotropic constants $K_1$ and $K_2$ satisfy the conditions for easy $c$-axis ($K_2$ $>$ -$K_1$) and basal-plane ($K_2$ $<$ -$K_1/2$) of magnetization for KReC and KOsC/KIrC, respectively. 

The experimentally and theoretically estimated Curie-Weiss temperatures are in good agreement with the theoretically estimated value of onsite Coulomb repulsion $U$ using linear response theory ($U$ $\approx$ 3 eV for KIrC and $U$ $\approx$ 7 eV for KReC). Interestingly, the antifluorites exhibit large magnetic anisotropy energy (MAE) $\approx$ 0.6-4 meV/transition metal which is one-to-two order higher than traditional MAE materials such as transition metals and multilayers formed out of them. Furthermore, we find that compressive/tensile strain can change the easy and hard axes as well as the magnetic ordering. Most importantly, strain enhances the MAE by one order and becomes giant for KOsC ($\approx$ 20-40 meV/Os) to make these antifluorites  promising applications in magnetic memory and storage devices. The second-order perturbative analysis of SOC provides good insight into the origin of large MAE and easy axis/plane of magnetization obtained from total energy density functional theory calculations.\\

\section{Acknowledgement}
This work was funded by the Department of Science and
Technology, India, through Grant No. CRG/2020/004330. We also acknowledge the use of computing resources at HPCE, IIT Madras.

\end{document}